\documentclass[11pt]{article}
\usepackage[dvips]{epsfig,color}
\usepackage[dvips]{graphicx, color}
\def\Re {\mbox{Re}\,}
\def\Im{\mbox{Im}\,}

\begin{document}

\title{Two-Particle Schr\"{o}dinger Equation Animations
of Wavepacket--Wavepacket Scattering (revised)}

\author{Jon J.V. Maestri\thanks{Present address: CH2M Hill, P.O.Box
428, Corvallis, OR 97330}, \, Rubin H. Landau\thanks{\tt
rubin@physics.orst.edu, http://www.physics.orst.edu/\~{\mbox{}}rubin}
\\ Oregon State University\\ Department of Physics\\ Corvallis, OR
97331\\ \and Manuel J. P\'{a}ez\\ Department of Physics\\ University
of Antioquia\\ Medellin, Colombia}

\maketitle

\begin{abstract}
A simple and explicit technique for the numerical solution of the
two-particle, time-dependent Schr\"{o}dinger equation is assembled and
tested.  The technique can handle interparticle potentials that are
arbitrary functions of the coordinates of each particle, arbitrary
initial and boundary conditions, and multi-dimensional
equations. Plots and animations are given here  and on the
World Wide Web of the scattering of two wavepackets in one dimension
\end{abstract}

\section{Introduction}

\begin{figure}
\begin{center}\label{schiff1}
\includegraphics[scale=.45]{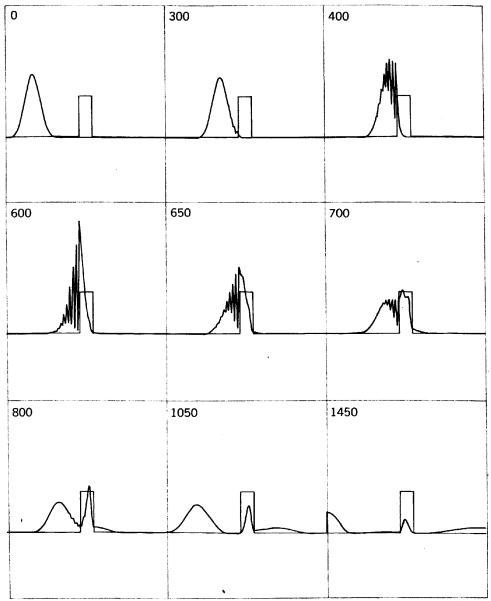}
\caption{A time sequence of a Gaussian wavepacket scattering from a
square barrier as taken from the textbook by Schiff. The mean energy
equals the barrier height.}
\end{center}
\end{figure}

Rather than showing the time dependence of two particles interacting
with each other, quantum mechanics textbooks often present a
time-independent view of a single particle interacting with an
external potential. In part, this makes the physics clearer, and in
part, this reflects the difficulty of solving the time-independent
two-particle Schr\"{o}dinger equation for the motion of wavepackets.
In the classic quantum mechanics text by Schiff \cite{schiff},
examples of realistic quantum scattering, such as that in
Fig.~\ref{schiff1}, are produced by computer simulations of wave
packets colliding with square potential barriers and wells.
Generations of students have carried memories of these images (or of
the film loops containing these frames \cite{goldberg}) as to what
realistic quantum scattering looks like.

While Fig.~\ref{schiff1} is a good visualization of a quantum
scattering processes, we wish to extend simulations of realistic
quantum interactions to include particle--particle scattering when
both particles are represented by wavepackets. Although more
complicated, this, presumably, is closer to nature and may illustrate
some physics not usually found in quantum mechanics textbooks. In
addition, our extension goes beyond the treatment found in most
computational physics texts which concentrate on {\em one-particle}
wavepackets \cite{koonin,gior,landau}, or highly restricted forms of
{\em two-particle} wavepackets \cite{qm}.

The simulations of the time-dependent Schr\"{o}dinger equation shown
by Schiff were based on the 1967 finite-difference algorithms
developed by Goldberg {\em et al.} \cite{goldberg}. Those simulations,
while revealing, had problems with stability and probability
conservation. A decade later, Cakmak and Askar \cite{cakmak} solved
the stability problem by using a better approximation for the time
derivative. After yet another decade, Visscher \cite{visscher} solved
the probability conservation problem by solving for the real and
imaginary parts of the wave function at slightly different
(``staggered'') times.

In this paper we combine the advances of the last 20 years and extend
them to the numerical solution of the {\em two particle}---in contrast
to the {\em one particle }---time-dependent Schr\"{o}dinger equation.
Other than being independent of spin, no assumptions are made
regarding the functional form of the interaction or initial
conditions, and, in particular, there is no requirement of separation
into relative and center-of-mass variables\cite{qm}. The method is
simple, explicit, robust, easy to modify, memory preserving, and may
have research applications. However, high precision does require small
time and space steps, and, consequently, long running times. A similar
approach for the time-dependent one-particle Schr\"{o}dinger equation
in a two-dimensional space has also been studied \cite{landau}.

\section{Two-Particle Schr\"{o}dinger Equation}

We solve the two-particle time-dependent Schr\"{o}dinger equation
\begin{eqnarray} \label{hpsi}
        i\frac{\partial}{\partial
        t} \psi(x_1,x_2,t) &=& H\psi(x_1, x_2,t)  ,\\
         H &=&  -\frac{1}{2m_1}\frac{\partial^2}{\partial x_1^2} -
        \frac{1}{2m_2}\frac{\partial^2}{\partial x_2^2} + V(x_1,x_2).\label{SE}
\end{eqnarray}
where, for simplicity, we assume a one-dimensional space and set
$\hbar =1$.  Here $H$ is the Hamiltonian operator and $m_i$ and $x_i$
are the mass and position of particle $i=1,2$. Knowledge of the
two-particle wave function $\psi(x_1,x_2,t)$ permits the calculation
of the probability density for particle 1 being at $x_1$ and particle
2 being at $x_2$ at time $t$:
\begin{equation}
        \rho(x_1,x_2,t) = \left|\psi(x_1,x_2,t)\right|^2.
\end{equation}
The fact that particles 1 and 2 must be located someplace in space leads to the
normalization constraint on the wave function:
\begin{equation}\label{Prob}
   \int_{-\infty}^{+\infty}
        \int_{-\infty}^{+\infty}  dx_1\, dx_2
        \left|\psi(x_1,x_2,t)\right|^2 = 1 .
\end{equation}

\begin{figure} \begin{center}
\includegraphics[angle=0,scale=.60]{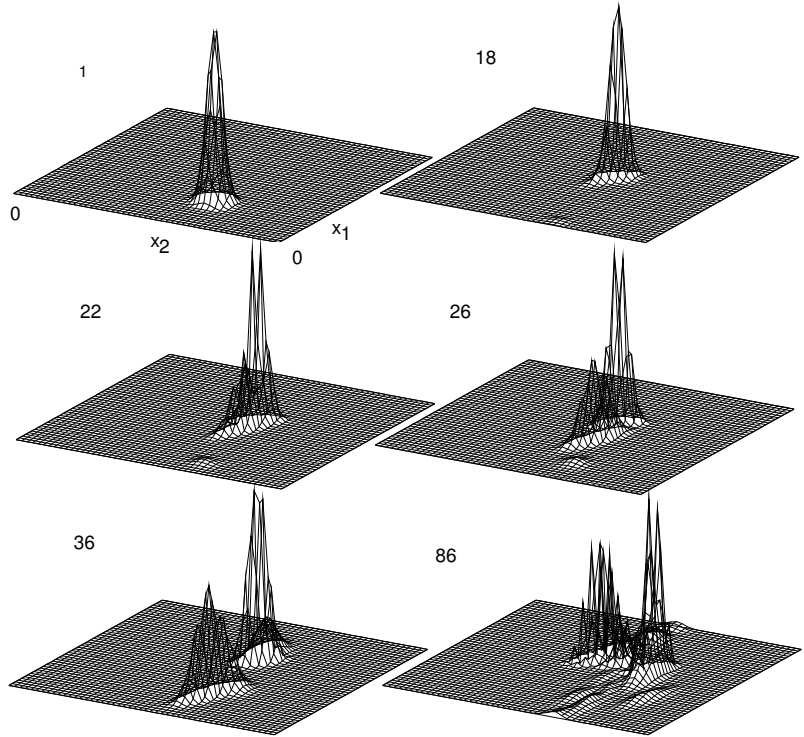}
\end{center} \caption{Six frames from an animation of the two-particle
density $\rho(x_1,x_2,t)$ as a function of the particle positions $x_1$
and $x_2$, for a repulsive $m$--10$m$ collision in which the mean
kinetic energy equals twice the barrier height. The numbers in the
left hand corners give the times in units of $100 \Delta
t$.\label{Rho}}
\end{figure}

The description of a single particle within a multi-particle system by
a single-particle wave function is an approximation unless the system
is uncorrelated (in which case the total wave function can be written
in product form).  However, it is possible to deduce meaningful
one-particle densities from the two-particle density by integrating
over the other particle:
\begin{equation}\label{rho1}
        \rho_1(x_i,t) = \int _{-\infty}^{+\infty} dx_j \,
        \rho(x_1,x_2,t), \ \ (i\ne j =1,2).
\end{equation}
Here we use a subscript on the single-particle density $\rho_i$ to
distinguish it from the two-particle density $\rho$. Of course, the
true solution is $\psi(x_1,x_2,t)$, but we find it hard to see the
physics in a three-variable complex function, and so, often, view
$\rho_1(x_1,t)$ and $\rho_2(x_2,t)$ as two separate wavepackets
colliding.

If particles 1 and 2 are identical, then their total wave function
should be symmetric or antisymmetric under interchange of the
particles. We  impose this condition on our numerical solution
$\psi(x_1,x_2)$, by forming the combinations
\begin{eqnarray}
        \psi^{'}(x_1,x_2) &=& {1\over\sqrt{2}}\left[\psi(x_1,x_2) \pm
\psi(x_2,x_1)\right]\ \ \ \  \Rightarrow \\
        2\rho(x_1,x_2) &=&
        \left|\psi(x_1,x_2)\right|^2 + \left|\psi(x_2,x_1)\right|^2
        \pm 2 \Re \left[\psi^*(x_1,x_2)\psi(x_2,x_1)\right].\label{symm}
\end{eqnarray}
The cross term in (\ref{symm}) places an additional correlation into
the wavepackets.

\section{Numerical Method}

We solve the two-particle Schr\"{o}dinger equation (\ref{hpsi}) via a
finite difference method that converts the partial differential
equation into a set of simultaneous, algebraic equations.  First, we
evaluate the dependent variable $\psi$ on a grid of discrete values
for the independent variables \cite{goldberg}:
\begin{equation}
        \psi(x_1,x_2,t) = \psi(x_1=l\Delta x_1,x_2
        =m\Delta x_2,t=n\Delta t)  \ \equiv \  \psi^n_{l,m} ,
\end{equation}
where $l$, $m$, and $n$ are integers.  The space part of the algorithm
is based on Taylor expansions of $\psi(x_1,x_2,t)$ in {\em both} the
$x_1$ and $x_2$ variables up to $\protect{\mathcal O}(\Delta x^4)$;
for example,
\begin{equation}
        \frac{\partial^2\psi}{\partial x_1^2} \simeq
        \frac{\psi(x_1+\Delta x_1,x_2) -2\psi(x_1,x_2) + \
        \psi(x_1-\Delta x_1,x_2)}{\Delta x_1^2} + \protect{\mathcal
        O}(\Delta x_1^2). \label{dxx1}
\end{equation}
In discrete notation, the RHS of the Schr\"{o}dinger equation
(\ref{hpsi}) now becomes:
\begin{equation}\label{SE2}
        H \psi =
        - \frac{\psi_{l+1,m}-2\psi_{l,m}+\psi_{l-1,m}}{2m_1\Delta x_1^2}
        - \frac{\psi_{l,m+1} - 2\psi_{l,m} + \psi_{l,m-1}}
        {2m_2\Delta x_2^2}  + V_{lm}\psi_{l,m}.
\end{equation}
Next, we express the time derivative in (\ref{hpsi}) in terms of
finite time differences by taking the formal solution to the
time-dependent Schr\"{o}dinger equation and making a
forward-difference approximation for time evolution operator:
\begin{equation} \label{growth}
        \psi_{l,m}^{n+1}  \ =\  e^{-i\Delta tH} \psi_{l,m}^n
        \ \simeq \ (1 -i \Delta t H ) \psi_{l,m}^n.
\end{equation}
Although simple, this approximation scheme is unstable since the term
multiplying $\psi$ has eigenvalue $(1-iE\Delta t)$ and modulus
$\sqrt{1+E^2\Delta t^2}$, and this means the modulus of the wave
function increases with each time step \cite{koonin}. The improvement
introduced by Askar and Cakmak \cite{cakmak} is a central difference
algorithm also based on the formal solution (\ref{growth}):
\begin{eqnarray}
        \psi^{n+1}_{l,m} - \psi^{n-1}_{l,m} &=& \left(e^{-i\Delta t H}
        - e^{i\Delta t H} \right)\psi^n_{l,m} \simeq -2i\Delta t H
        \psi^n_{l,m}, \\
        \Rightarrow \ \ \ \psi^{n+1}_{l,m} &\simeq& \psi^{n-1}_{l,m} -
        2i\left[\left\{(\frac{1}{m_1}+\frac{1}{m_2})4\lambda
        + \Delta x
        V_{l,m}\right\}\psi^n_{l,m} \right.  \label{algor}
        \\ && - \lambda
        \left\{\frac{1}{m_1}(\psi^n_{l+l,m} +
        \left.  \psi^n_{l-1,m}) + \frac{1}{m_2}(\psi^n_{l,m+1} +
        \psi^n_{l,m-1})\right\}\right] ,\nonumber
\end{eqnarray}
where we have assumed $\Delta x_1 =\Delta x_2$ and formed the ratio
$\lambda = \Delta t/\Delta x^2$.

Equation (\ref{algor}) is an {\em explicit} solution in which the wave
function at only two past time values must be stored simultaneously in
memory to determine all future times by continued iteration. In
contrast, an {\em implicit} solution determines the wave function for
all future times in just one step, yet this one step requires the
solution of simultaneous algebraic equations involving all space and
time values. Accordingly, an implicit solution requires the inversion
of exceedingly large $(\sim 10^{10} \times 10^{10}$) matrices.

While the explicit method (\ref{algor}) produces a solution which is
stable and second-order accurate in time, in practice, it does not
conserve probability  well.  Visscher\cite{visscher} has deduced
an improvement which takes advantage of the extra degree of freedom
provided by the complexity of the wave function to preserve
probability better. If we separate the wave function into real and
imaginary parts,

\begin{equation}
        \psi^{n+1}_{l,m} = u^{n+1}_{l,m} + i \, v^{n+1}_{l,m},
\end{equation}
the algorithm (\ref{algor}) separates into the pair of coupled equations:
\begin{eqnarray}        \label{psireal}
        u^{n+1}_{l,m} &=& u^{n-1}_{l,m} + 2\left[\left\{(\frac{1}{m_1} +
        \frac{1}{m_2})4\lambda + \Delta tV_{l,m}\right\}v^n_{l,m} \right.\\
        &&-
        \lambda\left\{\frac{1}{m_1}(v^n_{l+1,m}
        +
        \left.v^n_{l-1,m}) + \frac{1}{m_2}(v^n_{l,m+1} +
        v^n_{l,m-1})\right\}\right] , \nonumber \\
        \label{psiimg}
        v^{n+1}_{l,m} &=& v^{n-1}_{l,m} - 2\left[\left\{(\frac{1}{m_1}
        + \frac{1}{m_2})4\lambda + \Delta tV_{l,m}\right\}u^n_{l,m} \right.\\
        &&- \lambda\left.\left\{\frac{1}{m_1}   (u^n_{l+1,m}
         + u^n_{l-1,m}) \frac{1}{m_2}(u^n_{l,m+1}
        + u^n_{l,m-1})\right\}\right].\nonumber
\end{eqnarray}

\noindent
Visscher's advance evaluates the real and imaginary parts
of the wave function at slightly different (staggered) times,
\begin{equation}
          [u^n_{l,m},\, v^n_{l,m}]
        = [\Re\psi(x,t), \,
        \Im\psi(x,t+ {1\over 2}\Delta t)],
\end{equation}
and uses a definition for probability density that differs for integer
and half-integer time steps,
\begin{eqnarray}        \label{oneP2}
        \rho(x,t) &=& \left|\Re \psi(x,t)\right|^2
          + \Im \psi(x,t+\frac{\Delta t}{2}) \,
        \Im \psi(x,t-\frac{\Delta t}{2}), \\
        \rho(x,t+\frac{\Delta t}{2}) \label{halfP2}
        &=& \Re\psi(x,t+\Delta t)\Re\psi(x,t)
         + \left|\Im
        \psi(x,t+\frac{\Delta t}{2})\right|^2.
\end{eqnarray}
These definitions reduce to the standard one for infinitesimal $\Delta
t$, and provide an algebraic cancellation of errors so that
probability is conserved.

\begin{table}
\begin{center}
\begin{tabular}{ll}\hline
Parameter & Value\\ \hline $\Delta x$ & $0.001$\\ $\Delta t$ &
$2.5 \times 10^{-7}$\\ $k_1$ & $+157$ \\ $k_2$ & $-157$ \\ $\sigma $& $0.05$ \\
$x_{1}^0$ & $467$ ($0.33 \times 1401$ steps)\\ $x_{2}^0$ & $934$
($0.667 \times 1401$ steps)\\ $N_1=N_2$ & $1399$\\ $L$ & $1.401$ ($1401$
space steps) \\ $T$ & $0.005$ ($20,000$ time steps) \\ $V_0$ & $-100,000$\\
$\alpha$ & $0.062$\\ \hline
\end{tabular}
\caption{Table 1, Parameters for the antisymmetrized, m--m collision
with an attractive square well potential.\label{tab1}}
\end{center}
\end{table}

\section{Simulations}

We assume that the particle--particle potential is central and depends
only on the relative distance between particles 1 and 2 (the method
can handle any $x_1$ and $x_2$ functional dependences).  We have
investigated a ``soft'' potential with a Gaussian dependence, and a
``hard'' one with a square-well dependence, both with range $\alpha$
and depth $V_0$:
\begin{equation} \label{V}
        V(x_1,x_2) \ = \ \cases{V_0 \exp[-{|x_1-x_2|^2 \over 2 \alpha^2}]
         & (Gaussian) \cr
        V_0 \ \theta(\alpha - |x_1-x_2|)  & (Square) \cr }.
\end{equation}

\begin{figure} \begin{center}
\includegraphics[angle=0,scale=.45]{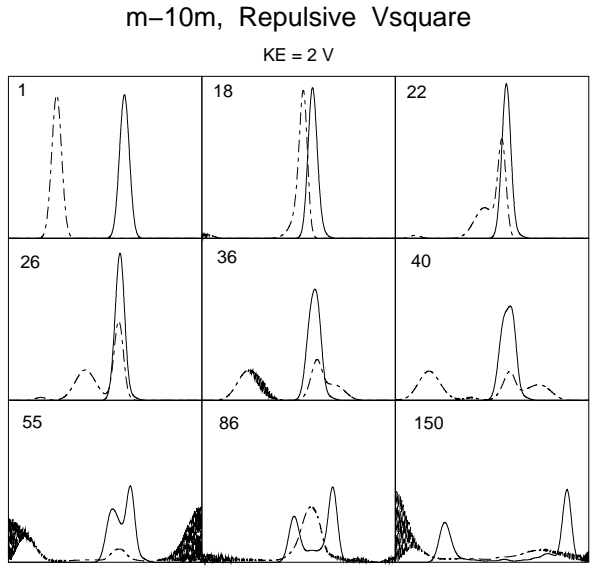}
\end{center} \caption{A time sequence of two Gaussian single-particle
wavepackets scattering from each other under the influence of a square
barrier. The mean energy equals twice the barrier height. The dashed
curve describes a particle of mass $m$ and the solid curve one of mass
$10m$. The number in the upper left-hand corner of each frame is the
time in units of $100 \Delta t$, and the edges of the frames
correspond to the walls of the box.
\label{m10msq9}} \end{figure}

\subsection{Initial and Boundary Conditions}

We model a scattering experiment in which particle 1, initially at
$x_{1}^0$ with momentum $k_{1}$, collides with particle 2, initially
far away at $x_{2}^0$ with momentum $k_{2}$, by assuming a product of
independent wavepackets for particles 1 and 2:
\begin{equation}
        \psi(x_1,x_2,t=0) = e^{ik_1x_1}\exp
        [{-\frac{(x_1-x_{1}^0)^2}{4\sigma^2}}] \times
        e^{ik_2x_2}\exp
        [{-\frac{(x_2-x_{2}^0)^2}{4\sigma^2}}].\label{initwf}
\end{equation}
Because of these Gaussian factors, $\psi$ is not an
eigenstate of the particle $i$ momentum operators $-i\partial/\partial
x_i$, but instead contains a spread of momenta about the mean, initial
momenta $k_{1}$ and $k_{2}$. If the wavepacket is made very broad
($\sigma\rightarrow \infty$), we would obtain momentum eigenstates.
Note, that while the Schr\"{o}dinger equation may separate into one
equation in the relative coordinate $x$ and another in the
center-of-mass coordinate $X$, the initial condition (\ref{initwf}),
or  more general ones, cannot be written as separate conditions on
$x$ and $X$.  Accordingly, a solution of the equation in each particle
coordinate is required \cite{qm}.

We start the staggered-time algorithm with the real part the wave
function (\ref{initwf}) at $t=0$ and the imaginary part at $t=\Delta
t/2$. The initial imaginary part follows by assuming that $\Delta t/2$
is small enough, and $\sigma$ large enough, for the initial time
dependence of the wavepacket to be that of the plane wave parts:
\begin{eqnarray}
               \Im\psi(x_1,x_2,t=\frac{\Delta t}{2}) &\simeq&
        \sin\left[k_1x_1+k_2x_2 - \left(\frac{k _{1}^2}{2m_1}
        +\frac{k_{2}^2}{2m_2}\right)\frac{\Delta t}{2}\right] \nonumber \\ &&
        \times \exp -\left[\frac{(x_1-x_{1}^0)^2 +
        (x_2-x_{2}^0)^2}{2\sigma^2}\right] .
\end{eqnarray}

In a scattering experiment, the projectile enters from infinity and
the scattered particles are observed at infinity. We model that by
solving our partial differential equation within a box of side $L$
(ideally) much larger than both the range of the potential and the
width of the initial wavepacket. This leads to the boundary conditions
\begin{equation} \label{bc}
        \psi(0,x_2,t)=\psi(x_1,0,t)=\psi(L,x_2,t)=\psi(x_1,L,t)=0.
\end{equation}
The largeness of the box minimizes the effects of the boundary
conditions during the collision of the wavepackets, although at large
times there will be interesting, yet artificial, collisions with the
box.

Some typical parameters used in our tests are given in
Table~\ref{tab1} (the code with sample files are available on the on
Web \cite{nacphy}). Our space step size $\Delta x = 0.001$ is
$1/1,400$th of the size of the box $L$, and $1/70$th of the size
($\sqrt{2} \sigma \simeq 0.07$) of the wavepacket. Our time step
$\Delta t = 2.5 \times 10^{-7}$ is $1/20,000$th of the total time $T$,
and $1/2,000$th of a typical time for the wavepacket
[$2\pi/(k_1^2/2m_1) \simeq 5 \times 10^{-4}$].  In all cases, the
potential and wavepacket parameters are chosen to be similar to those
used in the one-particle studies by Goldberg {\em et al.}.  The time
and space step sizes were determined by trial and error until values
were found which provided stability and precision (too large a $\Delta
x$ leads to spurious ripples during interactions).  In general,
stability is obtained by making $\Delta t$ small enough
\cite{visscher}, with simultaneous changes in $\Delta t$ and $\Delta
x$ made to keep $\lambda = \Delta t/\Delta x^2$ constant.  Total
probability, as determined by a double Simpson's-rule integration of
(\ref{Prob}), is typically conserved to 13 decimal places,
impressively close to machine precision. In contrast, the mean energy,
for which we do not use a definition optimized to staggered times, is
conserved only to 3 places.

\begin{figure} \begin{center}
\includegraphics[angle=0,scale=.45]{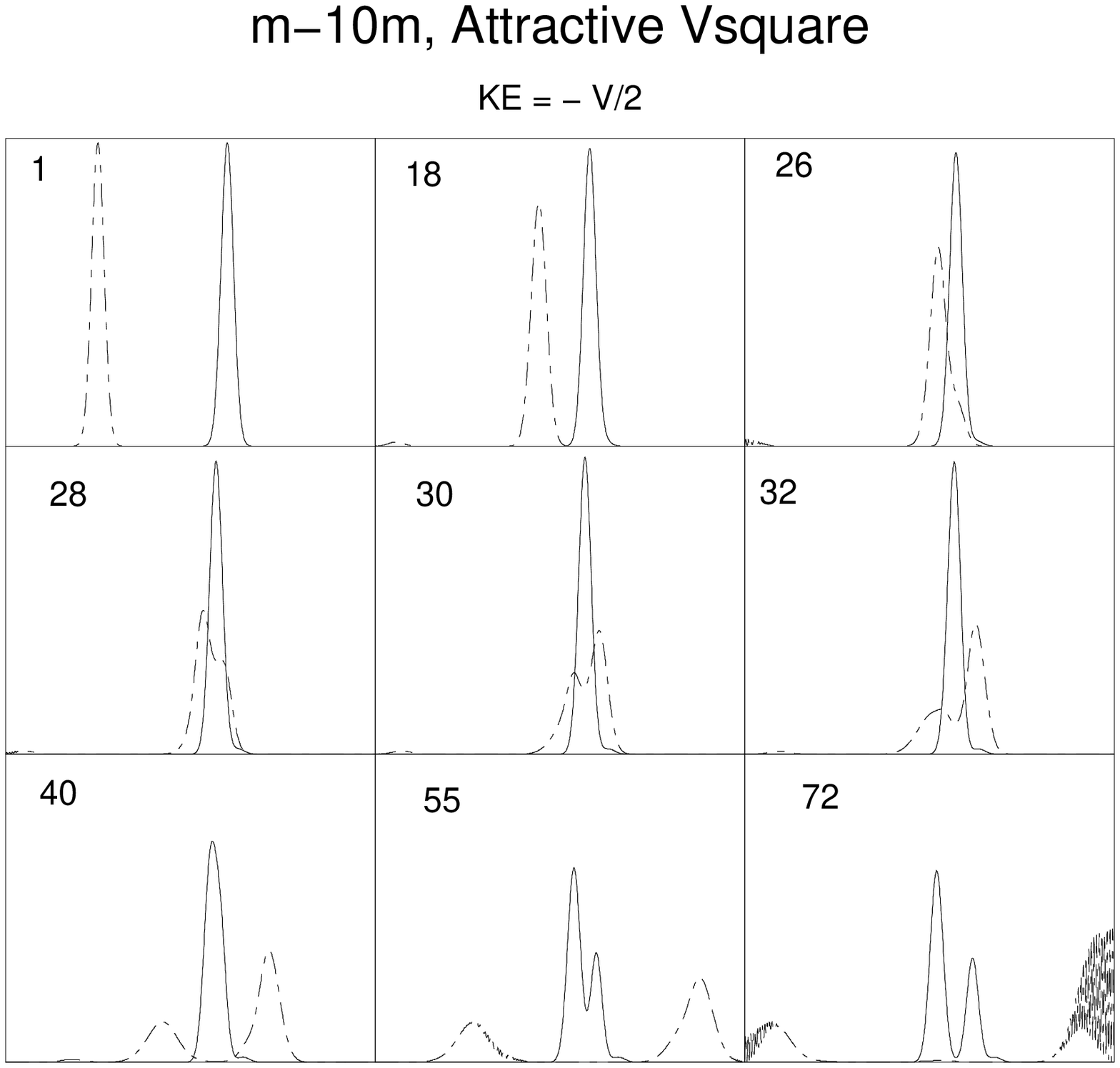}
\end{center} \caption{Same as Fig.~\ref{m10msq9} except now the
potential is attractive with the mean energy equal to half the depth.
\label{m10msq4}} \end{figure}

\subsection{Barrier-Like Collisions}

We solve our problem in the center-of-momentum system by taking $k_2 =
-k_1$ (particle 1 moving to larger $x$ values and particle 2 to
smaller $x$). Our first simulations and Web animations \cite{nacphy}
emulate the one-particle collisions with barriers and wells studied by
Goldberg {\em et al.} and presented by Schiff. We make particle 2 ten
times heavier than particle 1, so that particle 2's initial wavepacket
moves at $1/10$th the speed of particle 1's, and so looks like a
barrier. Although we shall describe several scattering events, the
animations available on the Web speak for themselves, and we recommend
their viewing.

In Fig.~\ref{Rho} we show six frames from an animation of the
two-particle density $\rho(x_1,x_2,t)$ as a simultaneous function of
the particle positions $x_1$ and $x_2$. In Fig.~\ref{m10msq9} we show,
for this same collision, the {\em single-particle} densities
$\rho_{1}(x=x_{1},t)$ and $\rho_{2}(x=x_{2},t)$ extracted from
$\rho(x_{1},x_{2},t)$ by integrating out the dependence on the other
particle via (\ref{rho1}). Since the mean energy equals twice the
maximum height of the potential barrier, we expect complete
penetration of the packets, and indeed, at time 18 we see that the
wavepackets have large overlap, with the repulsive interaction
``squeezing'' particle 2 (it gets narrower and taller). During times
22--40 we see part of wavepacket 1 reflecting off wavepacket 2 and
then moving back to smaller $x$ (the left). From times 26--55 we also
see that a major part of wavepacket 1 gets ``trapped'' inside of
wavepacket 2 and then leaks out rather slowly.

\begin{figure} \begin{center}
\includegraphics[angle=0,scale=.45]{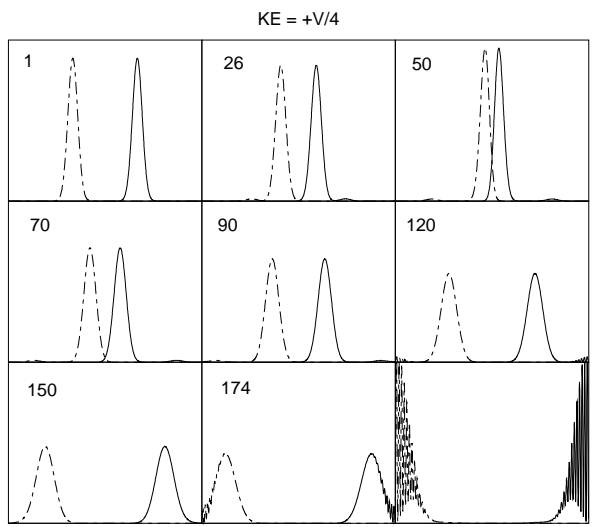}
\end{center} \caption[mmsq4]{Same as Fig.~\ref{m10msq4},
except now for a repulsive $m$--$m$ collision in which the mean energy
equals one quarter of the barrier's height. \label{mmsq4}}
\end{figure}

We see that for times 1--26, the $x_2$ position of the peak of
$\rho(x_{1},x_{2},t)$ in Fig.~\ref{Rho} changes very little with time,
which is to be expected since particle 2 is heavy. In contrast, the
$x_1$ dependence in $\rho(x_1,x_2,t)$ gets broader with time, develops
into two peaks at time 26, separates into two distinct parts by time
36, and then, at time 86 after reflecting off the walls, returns to
particle 2's position. We also notice in both these figures, that at
time 40 and thereafter, particle 2 (our ``barrier'' ) fissions and
there are two peaks in the $x_2$ dependence.

As this comparison of Figures \ref{Rho} and \ref{m10msq9}
demonstrates, it seems easier to understand the physics by
superimposing two single-particle densities (thereby discarding
information on correlations) than by examining the two-particle
density.  Accordingly, the figures we show hence, and the majority of
the animations on the Web, are of single-particle densities.

\begin{figure} \begin{center}
\includegraphics[angle=0,scale=.45]{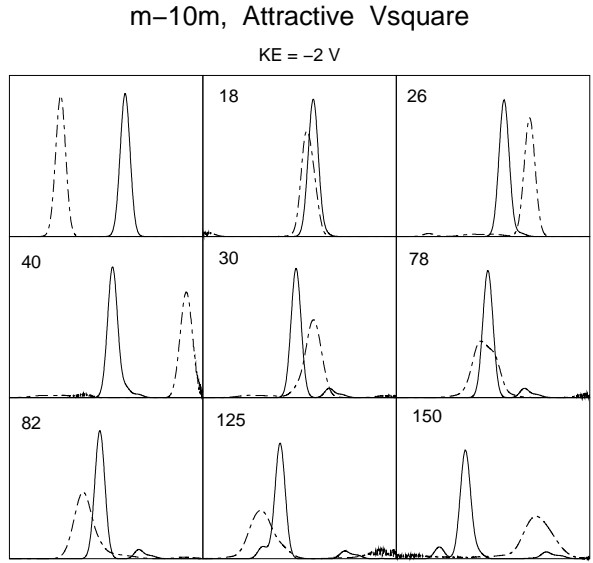}
\end{center} \caption[mmsq5]{Same as Fig.~\ref{m10msq4},
except now for an attractive $m$--$m$ collision in which the mean
energy equals one quarter of the well's depth. \label{mmsq5}}
\end{figure}

Fig.~\ref{m10msq9} is similar to the behavior present in Schiff's
one-particle simulation, Fig.~\ref{schiff1}, but without ripples
during the collision.  Those ripples are caused by interference
between scattered and incident wave, and even though we have a square
barrier potential acting between the particles, neither particle
``feels'' the discontinuity of the sharp potential edge at any one
time. However, there are ripples when our wavepackets hit the walls,
as seen at times 55 and 150. (There is also a ripple at time 36
arising from interference with the small, reflected part of the left
edge of 1's initial wavepacket. This artifact of having particle 1 so
near the left edge of the bounding box can be seen reflecting off the
left wall at time 18.)

Something new in Fig.~\ref{m10msq9}, that is not in Schiff, is the
delayed ``fission'' of the heavier particle's wavepacket after time 40
due to repulsion from the reflected and transmitted parts of
wavepacket 1. In addition, at time 86 we see that the reflected and
transmitted parts of 1 have reconstituted themselves into a single but
broadened wavepacket, that at time 150 is again being reflected from
the left wall.

In Fig.~\ref{m10msq4} we see another $m$--$10m$ collision. This time
there is an attractive interaction between the particles and again the
mean energy equals half the well depth. Even though the kinetic energy
is low, the interaction is attractive and so particle 1 passes through
particle 2. However, some of wavepacket 1 is reflected back to the left
after the collision, and, as we see at time 55, the wavepacket for the
heavy particle 2 fissions as a consequence of its attraction to the
two parts of wavepacket 1.

\begin{figure} \begin{center}
\includegraphics[angle=0,scale=.45]{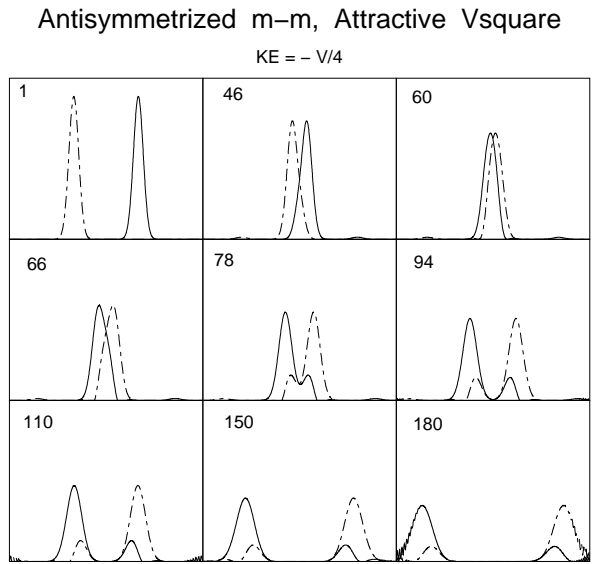}
\end{center} \caption[antisymm]{Same as Fig.~\ref{m10msq4},
except now for an attractive $m$--$m$ collision in which the mean
energy equals one quarter of the well's depth, and for which the
wavefunction has been antisymmetrized. \label{antisymm}}
\end{figure}

Although we do not show them here, on the Web we also display movies
of collisions corresponding to a Gaussian potential acting between
the particles. These are much softer collisions and have behaviors
similar to classical particles bouncing off each other, with squeezing
and broadening of the wavepackets, but little breakup or capture.

\subsection{$m$--$m$ Collisions}

In Fig.~\ref{mmsq4} we show nine frames from the movie of a repulsive
$m$--$m$ collision in which the mean kinetic energy equals one quarter
of the barrier height. The initial packets are seen to slow down as
they approach each other, with their mutual repulsion narrowing and
raising the packets up until the time (50) when they begin to bounce
back. The wavepackets at still later times are seen to retain their
shape, with a progressive broadening until they collide with the walls
and break up. As shown on the Web, when the mean energy is raised
there will be both transmitted and reflected wave, already seen in
Fig.~\ref{m10msq9} for an $m$--$10m$ collision.

In Fig.~\ref{mmsq5} we show nine frames from the movie of an
attractive $m$--$m$ collision in which the mean energy equals one
quarter of the well depth. The initial packets now speed up as they
approach each other, and at time 60 the centers have already passed
through each other. After that, a transmitted and reflected wave for
each packet is seen to develop (times 66--78). Finally, each packet
appears to capture or ``pickup'' a part of the other packet and move off
with it (times 110--180).

In Fig.~\ref{antisymm} we repeat the collision of Fig.~\ref{mmsq5},
only now for a wave function that has been {\em antisymmetrized}
according to (\ref{symm}). The anitsymmetrization is seen to introduce
an effective repulsion into what is otherwise an attraction (compare
the two figures for times 60--66). Again, some capture of the other
wavepacket is noted from times 94 on, only now the internal captured
wavepacket retains its Gaussian-like shape, apparently the result of
decreased interference.

\begin{figure} \begin{center}
\includegraphics[angle=0,scale=.45]{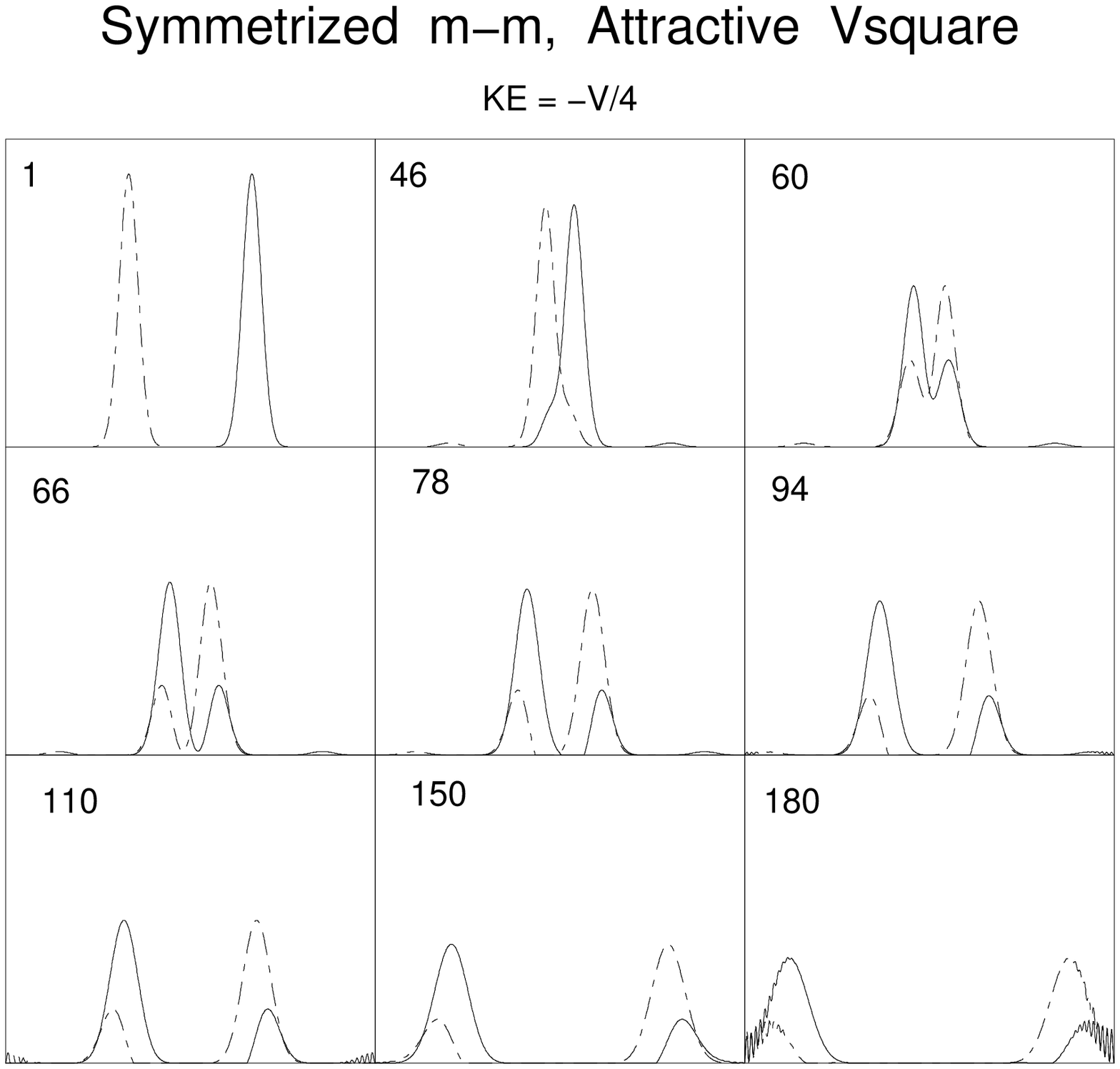}
\end{center} \caption[symm.ps]{Same as Fig.~\ref{m10msq4},
except now for an attractive $m$--$m$ collision in which the mean
energy equals one quarter of the depth, and for which the wavefunction
has been symmetrized. \label{symm.ps}}
\end{figure}

Finally, in Fig.~\ref{symm} we repeat the collisions of
Figures~\ref{mmsq5} and \ref{antisymm}, only now for a wave function
that has been {\em symmetrized} according to (\ref{symm}). The
symmetrization is seen to introduce an effective added attraction
(compare the three figures for time 60 which shows the greatest
penetration for the symmetrized case). While there is still capture of
the other wavepacket, the movie gives the clear impression that the
wavepackets interchange with each other as a consequence of the
symmetrization.

\section{Summary and Conclusions}

We have assembled and tested a general technique for the numerical
solution of the two-particle, time-dependent Schr\"{o}dinger equation.
Because the technique is general, application to two or three
dimensions and for other potentials and initial conditions should be
straightforward. For example, further studies may want to investigate
initial conditions corresponding to bound particles interacting with a
surface, or the formation of a molecule near a surface.

The Goldberg-Schiff's image (Fig.~\ref{schiff1}) of a wavepacket
interacting with a potential barrier is still a valuable model for
understanding the physics occuring during a particle's collision.
Here we have extended the level of realism to what a collision between
two particles looks like. In doing so with a simple square-well
potential between the two particles, we have discovered that fission
and particle pickup occur quite often, although the physics may be
quite different from that in nuclei.  While somewhat of a challenge to
understand fully, we have also provided animations of the behavior of
the two-particle density during collisions. We have placed the
animations, source codes, and movie-making instructions on the Web
with the hope that future students will also carry some of these
images of the quantum world with them.

\section{Acknowledgments}

We wish to thank an anonymous referee, Henri Jansen, Al Stetz, Al
Wasserman, and Roy Schult for helpful and illuminating
discussions. Support has been provided by the U.S. National Science
Foundation through the Northwest Alliance for Computational Science
(NACSE) and the REU program, and the U.S.  Department of Energy Office
of High Energy and Nuclear Physics. RHL wishes to thank the San Diego
Supercomputer Center and the Institute for Nonlinear Science at UCSD
for their hospitality.

\end{document}